\documentclass[10pt,twocolumn]{article}

\usepackage[top=1.0in, bottom=1.2in, left=1.0in, right=1.0in]{geometry} 

\usepackage{ifpdf}
\ifpdf   
\usepackage{thumbpdf} 
\RequirePackage[colorlinks,hyperindex,plainpages=false,pdfauthor={Robert Primmer},citecolor=black,urlcolor=blue,linkcolor=black]{hyperref}
\else
\RequirePackage[plainpages=true]{hyperref}
\usepackage{color}
\fi

\usepackage{wrapfig}     
\usepackage{subfig}    
\usepackage{graphicx}   
\usepackage{verbatim}   

\newcommand{\app}{application}
\newcommand{\apps}{applications}
\newcommand{\cmd}{custom metadata}
\newcommand{\Cmd}{Custom metadata}
\newcommand{\db}{database}
\newcommand{\dos}{distributed object stores}

\newcommand{\fs}{file system}
\newcommand{\fsm}{file system metadata}
\newcommand{\fss}{file systems}
\newcommand{\gdb}{graph database}

\newcommand{\hds}{Hitach Data Systems}
\newcommand{\hfs}{hierarchical file system}
\newcommand{\md}{metadata}
\newcommand{\os}{object store}
\newcommand{\oss}{object stores}

\newcommand{\RDOS}{relational distributed object store}

\newcommand{\sd}{structured data}
\newcommand{\ud}{unstructured data}

\makeatletter
\newcommand{\setversion}[1]{\def\@version{#1}}
\newcommand{\version}[0]{\@version}
\makeatother

\begin{document}

\setversion{1.0}  
\title{Creating a \RDOS{}}
\author{Robert Primmer, Scott Nyman, Wayzen Lin \\ \hds{} \\ \small \{bob.primmer,scott.nyman,wayzen.lin\}@hds.com}
\date{June 2013}

\maketitle

\ifpdf
\pdfbookmark[1]{Abstract}{abstract}
\fi

\begin{abstract}

In and of itself, data storage has apparent business utility. But when we can convert data to information, the utility of stored data increases dramatically. It is the layering of relation atop the data mass that is the engine for such conversion. Frank relation amongst discrete objects sporadically ingested is rare, making the process of synthesizing such relation all the more challenging, but the challenge must be met if we are ever to see an equivalent business value for \ud{} as we already have with \sd{}. This paper describes a novel construct, referred to as a \RDOS{} (RDOS), that seeks to solve the twin problems of how to persistently and reliably store petabytes of unstructured data while simultaneously creating and persisting relations amongst billions of objects. 

\end{abstract}

\section{Introduction} \label{s:intro}
Databases have proven to be a useful and versatile container for housing structured data, consisting of relatively simple but well-defined data types, referred to as \emph{scalars} (mostly numbers and strings).  In the 1970's databases evolved beyond a basic storage container to provide functions critical to the business by allowing relations amongst the stored data to be expressed and persisted \cite{Codd:1970:RMD:362384.362685}. This is fundamental to how business operates---where data from one part of the business must be correlated to other parts of the business if insights and efficiencies are to be realized. 

As a simple example, consider operations at a retailer such as Walmart where a database has two tables: one records point-of-sale transactions, the other holds inventory data. At the point-of-sale it's useful to have a record of all the items purchased, however it's far more useful to be able to then relate this back to the inventory system to automatically reorder goods as needed. This relation, and the automation it enabled, let Walmart eliminate manual shelf stock planning, lowered cost, and prevented over- and under-stocks. This basic example shows where adding relation provides significant efficiency to the business.

The container used to store such transactions is typically a relational \db{}, comprised of rows (tuples) and columns of scalar data. A substantial strength of relational databases comes in the form of a standard query and transformation logic, SQL, which allows applications outside the creating \app{} to query the data \cite{Angles:2008:SGD:1322432.1322433}. This ability to separate data from the bounds of the creating \app{} amplifies and extends the value the data can provide the business. 

\subsection{Database for Unstructured Data} \label{s:uddb}
In contrast to the simple scalar types typical to \sd{}, \ud{} is comprised of rich and expressive data types (e.g. PowerPoint presentations or full-motion video) that do not fit well into the traditional database paradigm. 

An \os{} is essentially a database for unstructured data, composed of two parts: a distributed database that holds \emph{object references} and a distributed file store that stores the user data, referred to as \emph{data objects} or \emph{blobs}.  The database is typically modeled as a NoSQL ``shared nothing'' data store for horizontal scaling---replicating and partitioning data over many servers \cite{Cattell:2011:SSN:1978915.1978919}. 

Public cloud examples include key-value stores such as Amazon's Dynamo \cite{DeCandia:2007:DAH:1323293.1294281} and Project Voldemort used by LinkedIn \cite{Sumbaly:2012:SLB:2208461.2208479}. In the enterprise and service provider sectors, the Hitachi Content Platform\footnote{\href{http://www.hds.com/products/file-and-content/content-platform}{http://www.hds.com/products/file-and-content/content-platform}}  (HCP) provides a distributed object store that typically resides behind a firewall \cite{RJP:2010:DOS-POP}. HCP is conceptually similar to the combination of Google's Bigtable \cite{Chang:2008:BDS:1365815.1365816} for storing object references and Google File System (GFS) \cite{Ghemawat:2003:GFS:945445.945450} for storing data objects.  

\subsection{Abstraction}
Fundamental to object storage is that the detail of the distributed database and underlying file systems are abstracted from both the client \apps{} (users) and system administrators. In the process \oss{} shift the client model, essentially presenting storage as a service rather than requiring clients to be directly involved in data storage decisions, such as properly balancing directory trees. While some \oss{} only support a single flat namespace \cite{RJP:2005:CA}, others allow the global namespace to be logically partitioned for greater security, e.g. with a collection of \emph{buckets} in Amazon S3 \cite{varia2008cloud}, or \emph{namespaces} in HCP \cite{MR:2013:HCP-ARCH}.

Such abstraction allows for comparatively na\"{\i}ve users and administrators as the \os{} takes care of the detail of \emph{how} and \emph{where} user data is stored, protected, geo-replicated,  de-duplicated, versioned, garbage collected, and so forth. This grossly simplifies application development and deployment, as evidenced by the plethora of start-ups who are able in short order to get a worldwide service up, running, and generating revenue by leveraging a public object storage service---something never seen before in history. 

Likewise, we see where limited compute platforms, such as smartphones and tablets, are able to have full access to a universe of data despite zero support for a single storage protocol. Such devices have no support for traditional storage protocols (FibreChannel, SCSI, NFS, CIFS) made popular in the last century for LAN-attached disk \cite{gibson1998cost}. Instead both simple and complex data types are served by nothing more than the basic web protocol HTTP. 

\subsection{Contributions}
This paper describes a mechanism for adding a \emph{relational layer} on top of the object storage layer, without destroying the simplicity gained through the abstractions for which \oss{} are noted. Our goal is for users to be able to manage relations between data in similar fashion to a relational database. However, since unstructured data is intrinsically different from structured data, it's necessary that we provide a substrate for defining and describing such relation. Further, we envision creating a mechanism by which users can query an RDOS in a manner similar to the way they query a relational database today for similar benefit. 

\subsection{Document Organization}
The remainder of the document is organized as follows. Section \ref{s:ud} defines \ud{} and touches on some of the associated challenges this datatype brings. Metadata is essential to the challenge of creating a relational layer on top of \ud{}. Section \ref{s:md} defines \md{}, explains why it is of such value, and identifies its potential to transform \ud{} from an undifferentiated data mass to a highly correlated data set that provides genuine value to the business. Section \ref{s:hcp} describes HCP in its present form and section \ref{s:rdos} describes the ideal of extending it to be a \RDOS{} and the business value it provides. Related work is described in section \ref{s:relwork}  and section \ref{s:conclusion} concludes with a summary of the topics covered in this paper.

\section{Unstructured Data} \label{s:ud}
The awkward term \emph{\ud} is intended to complement the term \emph{\sd}---which commonly refers to data stored in a database (DB) of some variety. Logically, \ud{} can be thought of as all data not stored in a DB, but most commonly it is used to refer to \emph{files} housed in a \hfs{}. Compared to scalars arranged by row and column within a DB, files can be far more expressive, comprising such varied formats as office documents, digital images, and full-motion video, collectively referred to as \emph{rich data types}.

\subsection{Objects} \label{s:obj}
The notion of files is further abstracted to \emph{objects}. While there isn't a canonical definition specifying precisely what constitutes an object in the context of storage, \os{} implementations tend to represent \emph{object} as the union of a data object, \fsm{} and (user created) \cmd{}. 

Here \emph{data object} is simply the user data; for example a Microsoft Word document that contains a travel itinerary. The \emph{\fsm{}} may include things such as the name of the file, the time it was created and when it was last updated. Thus far a basic \fs{} would suffice to house both the data (the Word document) and \fsm{} (the filename, time created, etc.). An \os{} provides the ability to add a third element, \emph{\cmd{}}, to the mix. 

\Cmd{} allows the user to annotate the base data with arbitrary text, often in the form of key-value pairs. Continuing our example, when storing our travel itinerary in an \os{} we may want to annotate this document with select key data such as: Year=2013, Department=Sales, Territory=US, Status=Approved. 

From this example we see that \cmd{} provides a mechanism to make the core file more useful by allowing the user to abbreviate the file with select information that can later be used to logically group documents. Additionally we can perform a lightweight search against the \os{} for all travel requests by Sales in 2013 that were approved for travel within the US and quickly return a list of all objects that meet these criteria. 

This is exactly the type of function we'd expect to perform against a sophisticated RDBMS. However, instead of being restricted to storing scalars, we gain the full expressiveness possible with rich data file types without sacrificing the ability to convert that data into information through mechanisms such as the ability to perform predicate searches. Better still, searching \md{} stored within the \os{} is a far lighter weight operation than searching the full content of every file stored and attempting to piece together logical groupings ad hoc. The value \md{} brings to \ud{} is substantial. In short, it's \md{} that allows us to add structure where none would otherwise exist in the \ud{} world. In \S\ref{s:md} we describe this value more deeply. 

\subsection{Data Growth}
Throughout the last two decades a recurring topic in IT is how to manage the exponential increase in the volume of data that must be stored as societies move increasingly to a digital world \cite{Gartner:2010:IT}. What's more, while the growth of data to date has been tremendous, the rate of increase is projected to grow greater still. IDC projects that the \emph{ Digital Universe} will grow by a factor of 300---to 40,000 exabytes!---by 2020, and that enterprises will have ``liability or responsibility for 80\% of the information'' in this digital universe. Further, \ud{} is expected to constitute 90\% of the 40,000 exabytes \cite{IDC:2012:DigitalUniverse}. 

This staggering growth correlates to a shift from text to rich data types such as digital images and video. To get a sense of \emph{how} such data growth can be rationalized, in \cite{RJP:2012:SDVSUD} we see where the capacity required to store a high definition movie trailer is 20,000 times greater than that required to store a traditional movie review. It's this continued shift to rich data types that fuels such spectacular growth; and this growth shows few signs of subsiding. 

\subsection{Data Containers}\label{s:dc}
Such growth begs the question: where are we to store this mountain of data? While RDBMS are used to hold \sd{}, \fss{} and \oss{} are common containers for unstructured data. 

Much of the R\&D over the last 50 years for \fss{} has focused on how to improve storage efficiency (e.g. optimizing \fs{} overhead), durability, and reliability. However, from the perspective of user interface, the basic paradigm of the \hfs{} remains largely unchanged since its introduction in the 1960s. 

For enterprise-class \oss{}, a cardinal focus has been on answering the question: How do we achieve extreme scale beyond that provided by traditional \fss{}? Since their genesis, the target market for \oss{} has been massive data stores, where the feature of cataloging all objects stored has greater value. 

It's comparatively easy to keep track of thousands of files for a few years; it's far more difficult to manage billions of them for decades. The Internet and the falling price of magnetic storage have shifted our expectations of digital storage. We now expect data to live in perpetuity, instead of being periodically culled for usefulness. This new behavior further highlights the need for large-scale data management both from a capacity and file count standpoint. 

A common misperception is that \oss{} are a replacement for \fss{}; instead, they are an augmentation. The \fs{} is tightly coupled to the operating system and provides a well established mechanism for organizing files within a hierarchy of directories. By contrast, an \os{} focuses on changing the presentation layer to the storage consumer through a simplified interface while achieving enormous scale by aggregating many \fss{} into a single, higher-order grouping. 

\begin{figure}[h]
  \centering
    \includegraphics[width=0.3\textwidth]{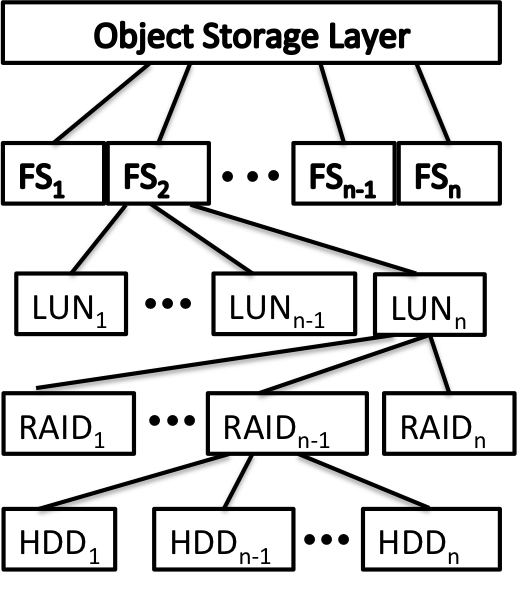}
  \caption{Storage Stack}
  \label{fig:stack}
\end{figure}

Figure \ref{fig:stack} presents an abstract view of the storage stack on a single node. The function of each superior layer is to aggregate and abstract the layer beneath, permitting greater sophistication and specialization in each layer without increasing complexity to clients of upper layers. The object storage layer creates a \emph{distributed storage service} to client \apps{} without requiring the clients to manage data distribution. 

\section{Metadata} \label{s:md}
In the past the term \emph{\md} was not widely known, understood mainly by technologist, and for good reason---\fss{} provide \md{} in parsimonious form, providing data about a file, such as when it was created and last updated. Object stores by contrast are comparatively lavish in what they allow for \md{}, allowing the user to associate any arbitrary text with a data object. 

Today, grandmothers think nothing of storing their photos online in globally available cloud repositories while annotating the pictures with  \md{} key-value pairs such as who is in the photo, where it was taken, and at what event. Young children know how to assign \md{} \emph{tags} to logically group items on Facebook and Twitter. In the 21st century, metadata has moved from the exotic to pedestrian. 

\subsection{Value of Metadata}
Metadata is the connective tissue that binds objects to one another. Additionally \md{} allows users to apply semantic meaning to otherwise opaque data objects, while providing a means to abbreviate large files with a very small amount of data. In short, it is \md{} that allows us to add structure to \ud{}.

How important is \md{}? In modern systems if the original designers fail to provide for it, the users do so themselves. Hashtags are a \md{} convention amongst users of the microblogging service Twitter \cite{Kwak:2010:TSN:1772690.1772751}, yet when it was first released Twitter had no support for \md{}. Instead hashtags were described by a user in 2007 (in a 140-character Twitter post), and became so popular that the engineers at Twitter added software support for it. Today, Twitter detects ``trending topics'' using popular hashtags \cite{Badia:2011:CAR:2070736.2070750}. 

\subsubsection{Logical Grouping}
Absent the ability to create logical groupings, a mass of objects housed in a data store is of limited value. In such a case the data store provides essentially equivalent value as tape---files are persisted, but there's not much you can do with them. 

Logical grouping creates tacit relation amongst an otherwise indistinct set of independent objects. Since this is a logical operation, the grouping is elastic and does not require expensive disk operations to move files into specific containers to define groupings. 

\subsubsection{Semantic Meaning} \label{s:sm}
Storing data is one thing, having that same data have genuine meaning to the user is quite another. For example, a photo stored on disk is valuable, but it's the ability to add semantic meaning to these photos (e.g., who is in the photograph at what event) that brings life to the data.  

\subsubsection{Index Efficiency}
Indexing the content of very large files consumes substantial capacity to hold the resulting index. Through \md{} it's possible to provide a comparatively small subset of data, effectively creating an abbreviation of the larger file content. The capacity required to index this \md{} can be orders of magnitude less than that required to index the full set of data objects, with a commensurate reduction in computes required to create the search index.

\section{HCP} \label{s:hcp}
Hitachi Content Platform (HCP) is a multipurpose distributed \os{} designed to support large-scale repositories of unstructured data. Physically, HCP is a collection of storage servers (nodes), referred to as a \emph{cluster}, that use a private back-end network for inter-node communications and a public front-end network for client communication. A gossip protocol is used determine active cluster membership. In the event of temporal failure of any node or service requests are automatically vectored to active nodes for fulfillment. 

The system is logically divided into a collection of \emph{tenants} and \emph{namespaces}; tenants are the administrative unit, namespaces are the storage unit. Tenants are critical for cloud / service provider deployments where it's important not only to logically separate client data, but also to divide the administration of each virtual instance of HCP. The system is designed so that no one administrator has dominion over the system as a whole, but only over their assigned tenant. A tenant administrator can create a collection of namespaces that will hold user data. 

The value proposition of a multitenant system is largely the same as any shared pool of physical resources such as SANs: Sharing reduces cost by pooling physical resources. In the process they also expose data on shared storage by unauthorized users and overwrites by multiple clients \cite{yoshida1999lun}. To counter this HCP employs a complex of security measures. These include strictly dividing administrative functions, defined user access rights and restrictions, access controls granular to individual objects, and all data is optionally encrypted, both in-flight and on disk.

Client access to the system is via a number of software gateways, distinguished by protocol. Presently the following protocols are supported: HTTP (REST), S3, CIFS, NFS, SMTP and WebDAV. Objects that were ingested using any protocol are immediately accessible through any other supported protocol. These protocols can be used to access the data with a web browser, HCP client tools, third-party applications, Windows Explorer, or native Windows or Unix tools \cite{MR:2013:HCP-ARCH}.

HCP provides high availability with strong consistency guarantees. Replication comes in two forms: intra- and inter-cluster. Intra-cluster replication is synchronous to the write path to ensure that all data ingested is successfully persisted to disk before returning acknowledgement to the client. Consistent hashing is used to distribute data among the nodes. Within the strictures of assuring replicas are assigned separate fault domains, writes bias toward lightly loaded nodes. Inter-cluster replication is asynchronous to the write path to ensure low write latency while providing geographic dispersion of objects. HCP uses XML\footnote{{\href{http://www.w3.org/XML}{http://www.w3.org/XML}}} as the serialization format for persisting \cmd{} and XPATH\footnote{\href{http://www.w3.org/TR/xpath}{http://www.w3.org/TR/xpath}} structured queries against that \md{}. This allows the system to return \emph{specific} answers to user queries, rather than a collection of \emph{potential} matches.

HCP employs redundancy at multiple levels: the object reference database, user data (data objects), and system and custom (user) \md{}. Likewise, all hardware components are redundant so there is no single point of failure. Commonly the focus for data protection centers on protecting user data. However, for database systems it's actually the protection of the index that's most important, as a loss of the pointer to the data is tantamount to the loss of the data itself. HCP shards the object reference database amongst distinct fault domains, both for performance and protection. Further, in the event of catastrophic failure, a \emph{scavenging service} is employed to reconstruct the \db{} from the constituent \md{} persisted to disk as part of the write path. 

The data model is one of immutable data objects with mutable \md{}. To simulate in-place updates of data objects HCP supports object versioning, i.e. the capability of a namespace to create, store, and manage multiple versions of objects within the repository. Capacity efficiency is achieved through a combination of compression and duplicate elimination. 

In addition to the actual data persistence component, the system needs to have scalable and robust solutions for load balancing, membership and failure detection, content verification, version control, retention policies, disposition of objects no longer under retention, compression, encryption, failure recovery, replica synchronization, overload handling, state transfer, garbage collection, concurrency and job scheduling, request marshaling, request routing, system monitoring and alarming, and configuration management.

The result is a system that provides data management as a service, referred to as \emph{Storage as a Service}, in a manner that makes both client development and management of the system simple. The design goal is that neither clients nor system administrators should be aware of the detail of the multiple software services needed to keep a multi-petabyte repository housing billions of objects coherent and responsive to client requests.  

\section{RDOS} \label{s:rdos}
The challenges of creating a relational \os{} are multiple and substantial. The first deals with the most basic question of how do we obtain needed \md{} in the first place (\S\ref{s:om}). To solve this we need to begin with a definition of what is needed (\S\ref{s:def}). We can then move on to the problem of how to extract and affine this \md{} to the associated data object (\S\ref{s:mxe}), how to create relations (\S\ref{s:cr}), how to persist these relations (\S\ref{s:pr}), and finally to put this information to good use through modeling and analytics (\S\ref{s:models}).

\subsection{Obtaining Metadata} \label{s:om}
To date there is relatively little metadata being associated with the data objects in object stores.  There are a number of reasons for this. 

\textbf{Application Reluctance}: Legacy \apps{} were built before a time when there was a well-defined means for storing and associating \md{} with data objects. However, even in the presence of such mechanisms today, a paucity of \md{} persists. First, a general \app{} by itself is unlikely to know what constitutes salient \md{} for a given file. Second, most commercial \apps{}, recognizing the value of \md{} to the customer, are reluctant to give up control of the \md{} to be used outside the confines of the application itself. Commercial \app{} providers may not be motivated to put extensive \md{} into a data store as it allows the customer to detach the data from the \app{}, in the process reducing the stickiness of the \app{}. This creates a financial disincentive to allow the user free and ready access to this \md{} without the need to involve the creating \app{}. 

Further, even if a commercial \app{} provider is properly motivated to do so for the good of the customer, it's hard to determine what is the right set of data to include in \md{} as each customer is different. Just as with a RDBMS, where individual customers create relations amongst data sets as appropriate to their own needs, the customer needs to be able to define the \md{} that is apropos to their specific environment. 

\textbf{Tools for Association}: Few tools exists to enable the systematic association of metadata to data objects.  Business agents with the necessary domain expertise for determining what constitutes relevant \md{} are unlikely to have the technical expertise required for tool development. 

\textbf{Sophisticated Metadata Stores}: Even when modern \apps{} wish to associate meaningful \md{} with data objects, few \oss{} have the sophistication to provide the necessary foundation for good programming practice. 

\begin{figure}[h]
  \centering
    \includegraphics[width=0.25\textwidth]{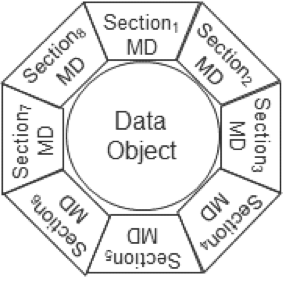}
  \caption{Object with Metadata Partitions}
  \label{fig:object}
\end{figure}

For example, for \md{} to be a first-class entity it's necessary that we are able to partition the address space so that different users can create, modify and delete their own \md{} in isolation. Today, most systems provide a single bucket where all \md{} is housed. Changes by one user therefore affect every other user, often in unpredictable ways. To protect user data from such unintended manipulations, all data stores partition the address space for data objects; few provide equivalent functionality for \md{}. For these reasons HCP allows \md{} partitions for each data abject. Figure \ref{fig:object} illustrates this concept with a data object that has eight distinct \md{} partitions. In \S\ref{s:mxe} we prescribe a mechanism for users to populate these \md{} partitions for both new data ingest, as well as for existing objects already stored.  

Finally, it's critical that an \os{} can scale---not just in its ability to store petabytes of data, but to be able to keep track of billions of discrete objects and associated \md{}. The former is comparatively easy, the latter is far more difficult. With HCP, each node can independently manage in excess of 800 million data objects and associated \md{}, a full cluster can manage 64 billion.

\subsection{Defining} \label{s:def}
The first step is to define what \md{} should be extracted or applied to data objects. Search engines can crack many file types and extract not only data from the core file, but also in some cases \md{} embedded in the file's header (e.g. in the case of DICOM images). However, standard data is is made far more useful by applying \emph{semantic meaning} to the content. This step is difficult for a general search engine to do as, by its very nature, semantic meaning is often particular to a user or organization and therefore doesn't lend itself to a generic solution.

Creating an ontology, where semantic meaning can be mapped to data, is referred to as a \emph{Data Dictionary}. Multiple such dictionaries can be created, each with meaning peculiar to a function within an organization. For example, one dictionary may contain a grammar common to a particular vertical, such as the field of healthcare, while other dictionaries will be particular to the organization itself, such as codenames used within a company. The sum of the individual data dictionaries creates the compendium necessary to inform the extraction and application step. 

\subsection{Extraction / Application} \label{s:mxe}
The data dictionaries defined in the prior step act as \emph{extractors} or \emph{applicators} run across a field of previously ingested data objects or against individual objects during ingestion. When applied as a filter, objects are scanned for relevant key-value pairs within the data object with the results applied as \md{} scalars. Of course, such extraction is only possible where the data object is searchable. In cases where it is not, such as for image files, applicators would apply \md{} as defined in the data dictionaries to relevant data objects. 

Since there can be multiple dictionaries, an object $O$ is run through a pipeline of size $p$ where the rules of each dictionary $d$ are applied in turn. For extraction operations we have $f_1 = \sum_{i=1}^p O \cap d_i$, for applicator operations we have $f_2 = \sum_{i=1}^p O \cup d_i$. In cases where a particular step in the pipeline has no rules to apply against an object that step becomes the identity function, that is $O_{input} = O_{output}$. Note that while it's possible that every step in a pipeline will be exclusively of type $f_1$ or $f_2$, this is not requirement; that is, $f_1$ and $f_2$ are not mutually exclusive. 

Once run through the pipeline, data objects are coupled with relevant \md{} and stored as scalars, which can be used for subsequent queries by client \apps{} and to inform relations between objects.

\begin{figure}[h]
  \centering
    \includegraphics[width=0.5\textwidth]{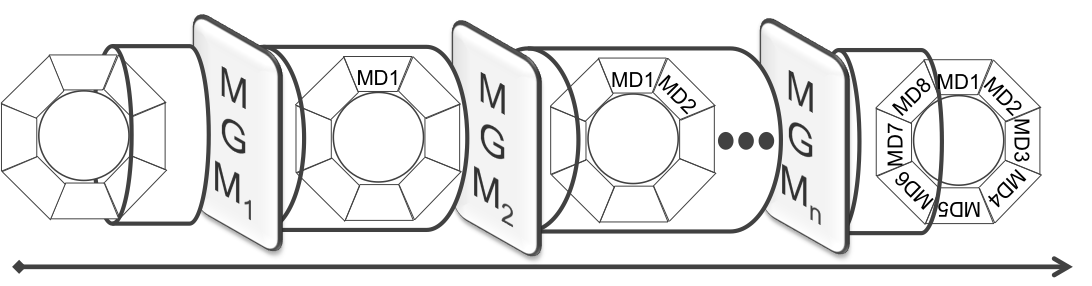}
  \caption{Metadata Generation Pipeline}
  \label{fig:pipe}
\end{figure}

In this model the extraction or application steps in the pipeline are referred to as a \emph{\md{} generation module}, or MGM. As depicted in Figure \ref{fig:pipe}, a data object progresses through a series of MGMs, where an MGM comprises a script language specific to the task. At every stage a decision is made whether there is \md{} to be applied. While each MGM is independent, the ordering of MGMs can matter as the algorithm applied at any individual MGM may base a decision of whether to add, change or delete a \md{} section based not only the data object itself, but rather the union of the data object and \md{} accumulated in the pipeline to that point. Nonetheless, fundamental to this design is that an MGM is independent, to allow the flexibility of adding or updating MGMs in existing pipelines.

\begin{figure}[h]
  \centering
    \includegraphics[width=0.5\textwidth]{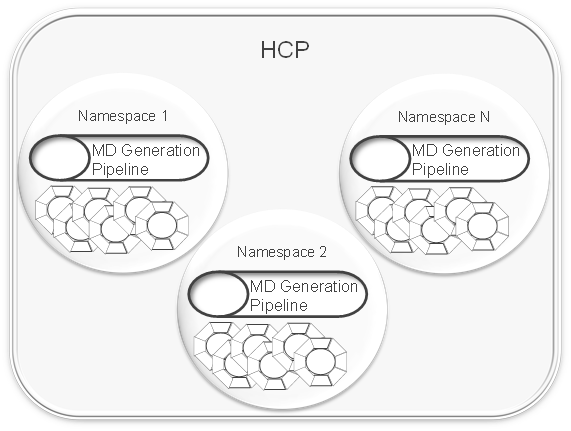}
  \caption{Unique Pipeline per Namespace}
  \label{fig:mgp}
\end{figure}

In HCP the address space is partitioned into \emph{namespaces} that essentially act as a virtual instance of the system as a whole. This segregation allows different data management and access policies to be applied to each set of data objects housed within a particular namespace.  It is a natural extension that each namespace may have a different set of MGMs to analyze data objects specific to a namespace (Figure \ref{fig:mgp}). This allows finer-grain decisions to applied to objects already grouped by user.

Database programmers likely recognize that the MGM construct is conceptually similar to a \emph{stored procedure} in a RDBMS. A stored procedure is a subroutine stored in the \db{} data dictionary that runs on the \db{} server itself rather than directly on the client. In this model the MGM acts as a stored procedure, where common code can be run directly on the RDOS cluster rather than the client. 

Note that a client \app{} could choose to perform all the functions of the MGM pipeline directly by reading an object, applying \md{}, and then rewriting this object back to the RDOS cluster. The downside of this method is that network bandwidth is consumed by the roundtrip for the read/write operations. It can therefore be more efficient to perform this same function on the cluster by running objects through the MGM pipeline. MGM scripts can be created either by the \app{} provider or constructed by the system administrator through a GUI. 

\subsection{Creating Relations} \label{s:cr}
Armed with the \md{} from \S\ref{s:mxe} we move to the next step of using this \md{} to create relations among objects. For this discussion we use the example of an auto insurance company servicing a customer claim after a vehicle collision. We have three components to the claim: the claim form, digital photographs taken by the appraiser of the damage sustained, and a scan of the police report. These rich data types are good candidates for an \os{} as all are \ud{}. 

Traditional \apps{} create a \db{} to collect transactions and create a schema to associate\db{} tables with one another. However, mobile \apps{} may not wish to require a local \db{}, instead choosing a remote \db{} for this function. In \S\ref{s:uddb} we state that \oss{} are a \db{} for \ud{}. Here RDOS acts as the \db{} for both the scalar data (in the form of key-value \md{} tags) and the data objects (blobs), which consist of any arbitrary file type. To use RDOS in our insurance example we begin with the claim form where a Claim ID (\texttt{CID}) is assigned (\texttt{CID=1234}) and ingested into RDOS. Subsequently the photos from the appraiser and the scan of the police report are uploaded. For all objects there is a \md{} tag set where \texttt{CID=1234}, providing a common key amongst the objects. At ingest, the client \app{} can choose to add a second \md{} field indicating that the images are related to the claim form. As objects are identified by URI, this translates to: \texttt{RelTo=URI\{Object1\}}. Through this structure we have a means of relating the full set of objects which are all associated with the same insurance claim. 

This example depends upon the \app{} explicitly creating the relation between the objects by setting the \texttt{RelTo} \md{} tag, which is the preferred method for new object ingest. However, the same mechanism described in \S\ref{s:mxe} for adding \md{} to objects already stored can be applied here. The final stage of a MGM pipeline can be used to assign the relationship tag to objects in the same manner as all other \md{}. The difference in this final MGM script is that it must persist all \md{} tags found when enumerating over a set of existing objects. The system administrator can then define how these objects are to be related via a GUI, which in turn creates the needed MGM script.

\subsection{Persisting Relations} \label{s:pr}
Once we've done the hard work of extracting key \md{} and presenting mechanisms to the user (both human and application) to define relations, we need to select a means of persisting these relations in a manner that makes sense for the data type. In our model we have chosen a \gdb{} for this function as the graph model provides an excellent means of describing relation. Graph database models can be defined as those in which data structures for the schema and instances are modeled as graphs or generalizations of them, and data manipulation is expressed by graph-oriented operations and type constructors \cite{Angles:2008:SGD:1322432.1322433}. 

Graph databases are described as ``whiteboard friendly.'' Drawing circles and connecting them with lines on a whiteboard is how we visualize graphs. We denote data points as \emph{nodes} and connect (relate) nodes with \emph{links}\footnote{Graph theory is a branch of discrete mathematics. Nodes are referred to as \emph{vertices} and what we call links are referred to as \emph{edges}, where edges connect pairs of distinct vertices. A simple graph (such as we'll be using in our descriptions) with $V$ vertices has at most $V(V-1)/2$ edges. We won't be dealing with graphs mathematically other than to note that a graph is defined by its vertices and its edges, not by the way we choose to draw it \cite{DBLP:books/daglib/0001826}.}, as illustrated in Figure \ref{fig:epi} (p. \pageref{fig:epi}). Graph databases, such as NEO4j \cite{Webber:2012:PIN:2384716.2384777}, are well suited to \ud{} as they are typeless and have no set schema, while providing several advantages for our application. 

First, the graph data model is useful when the interconnectivity of data and the ability to discover the \emph{relationships between values} is important, rather than simply commonality \emph{among value sets} typical to relational models. Discovering relations is fundamental to our design, so its important we select a data structure that can readily describe relations with good performance. Unlike join operations in relational databases or map-reduce operations in other databases, graph traversals are in constant time \cite{redmond2012seven}. Graph databases make it easy to discover centrality, where we measure individual nodes against a full graph. (The most famous centrality algorithm is Google PageRank \cite{ilprints422}. In \S\ref{s:models} we use centrality in an epidemiology example.) Finally, graph databases have been shown to perform full-text character searches significantly faster than relational databases \cite{Vicknair:2010:CGD:1900008.1900067}.  

Popular examples of publicly available \apps{} that make use of graph databases include Freebase\footnote{\href{http://www.freebase.com}{http://www.freebase.com}} and the Disease Ontology database\footnote{\href{http://disease-ontology.org}{http://disease-ontology.org}}. Freebase, created by Metaweb and acquired by Google in 2010, is a large community \md{} \db{} described as ``an open shared database of the world's knowledge.'' Disease Ontology, created by the University of Maryland School of Medicine, represents a comprehensive knowledge base of 8043 inherited, developmental and acquired human diseases \cite{schriml2012disease}.

\subsection{Models and Analytics} \label{s:models}
Once the data is available and programmatically accessible through well-defined APIs, modeling types that users can perform can be considered. 

Object associations can be definitive or manufactured through a probabilistic generative model to guide inference from incomplete data.  The former can be prescribed by the user by linking objects through a GUI or template; the latter through Bayesian inference \cite{Pearl:1988:PRI:52121}, which provides a rational framework for updating beliefs about latent variables in generative models given observed data \cite{MacKay:2002:ITI:971143}. 

Creating Bayesian models through graphs and predicate logic is more commonly the domain of data scientists than IT users. However, the goal for our system isn't to mandate a specific model of relation (definitive or derived), but rather to take care in design to not implicitly restrict the user to a particular model. Our system must provide the flexibility to persist and mutate object relation to meet a variety of user requirements\footnote{Nate Silver discusses the breadth of problems to which Bayesian reasoning can be applied in \cite{silver2012signal}.}. 

As such, our goal is to provide an abstract framework over which schemas can be applied with sufficient flexibility to allow these relations to span the prosaic, e.g. linking monotonically increasing instances of an object (creating a version tree), to the expressive, e.g. a graph schema where nodes represent variables and directed edges between nodes represent probabilistic causal links. 

\begin{figure}[h]
  \centering
    \includegraphics[width=0.3\textwidth]{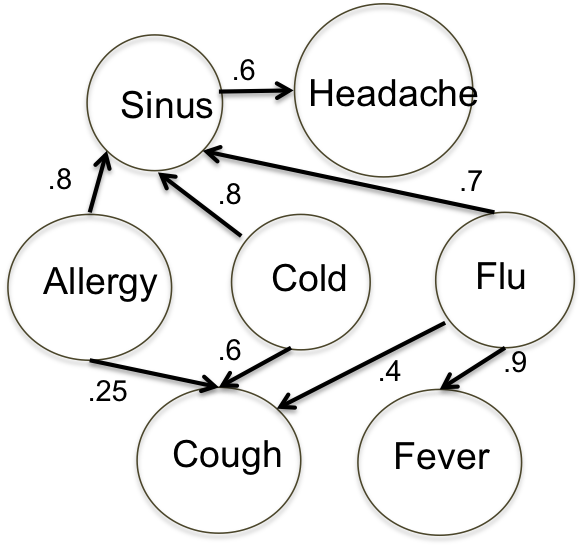}
  \caption{Epidemiology Directed Graph}
  \label{fig:epi}
\end{figure}

For example, in an epidemiology study nodes might represent whether a patient has a cold, a cough, a fever or other conditions, and the presence or absence of links indicates that colds tend to cause coughing and sinus inflammation but not fever; sinus inflammation tends to cause headache but not fever; and so on \cite{tenenbaum2011grow}. The probability of a causal relation can be further refined by applying a weighting (0 - 1.0) to each link; e.g. there's a 60\% likelihood that a cold will result in a cough. As depicted in Figure \ref{fig:epi} graphs provide a ready means for users to interpret data and visualize relationships. 

Through the proper application of multi-predicate constraints over the field of objects in a domain, the user is able to resolve to \emph{classes} of objects---such as all objects associated with the progenitor of a disease. For example a user might want to view all outcomes for patients over 60 in Africa with diminished respiratory function in the presence of the bacteria Legionella, and contrast this to those in the presence of the virus adenovirus to determine whether better outcomes are seen in patients with bacterial versus viral pneumonia. 

Such analytic questions can be compute intensive when run over the full mass of \ud{} (what we refer to as \emph{data objects} in \S\ref{s:obj}), but are comparatively lightweight operations when run against the much smaller set of \md{} associated with these data objects. Here the \md{} acts an abbreviation of the objects and allows us to create information from an otherwise undistinguished data mass.

\section{Related Work} \label{s:relwork}

The NoSQL space has been quite active, with marked similarity among implementations. Distributed systems can be distinguished by how they choose to bias availability versus consistency. Brewer's CAP theorem \cite{Gilbert:2002:BCF:564585.564601} states that a distributed system can provide for any two of the three attributes of \emph{Consistency, Availability} and \emph{Partition tolerance}. Since networks (particularly WANs) will always partition (i.e., over time there will be some parts of the network that are temporarily unreachable), for distributed systems this distills to a choice of a design that favors strong consistency or one that is highly available in the face of partitions. 

This is the heart of the distinction between SQL and NoSQL systems. SQL systems adhere to ACID properties while most NoSQL stores follow BASE properties\footnote{ACID stands for Atomicity, Consistency, Isolation and Durability; BASE stands for Basically Available, Soft state, Eventual consistency. Descriptions of ACID and BASE can be found in \cite{Wright:2007:EAS:1242520.1242521} and \cite{Cattell:2011:SSN:1978915.1978919} respectively.}. Many NoSQL implementations provide \emph{eventual consistency} \cite{Vogels:2008:EC:1466443.1466448}, where writes are permitted in the presence of partitions. In such systems a client can update any replica of an object and all updates to an object will eventually be applied, but potentially in different orders at different replicas, thus creating temporal inconsistencies that must be sorted out by client \apps{} \cite{Cooper:2008:PYH:1454159.1454167}. 

Amazon's Dynamo \cite{DeCandia:2007:DAH:1323293.1294281} is the classic example of the eventually consistent model. Dynamo, a key-value store, is a zero-hop distributed hash table, where each node maintains enough routing information locally to route a request to the appropriate node directly. Dynamo shares a number of characteristics with other \dos{} used by public cloud vendors that stem from their operational model, which is a closed loop system where both clients and all services are under the control of a single company. These systems, while used to support public facing services, can assume that their \emph{internal} operating environment is non-hostile and therefore have few security requirements such as authentication and authorization. Further, since the NoSQL \oss{} permit inconsistency, users (client \apps{}) must contain an agreed understanding of how to resolve conflicts during reads. As a general-purpose system with non-specific clients, HCP can make neither assumption and must design for hostile environments and provide strong consistency guarantees. 

Other systems that use  an eventual consistency model include Facebook's Cassandra (now Apache) and Google's Bigtable.  Bigtable is a distributed storage system for managing structured data. It maintains a sparse, multi-dimensional sorted map and allows applications to access their data using multiple attributes \cite{Chang:2008:BDS:1365815.1365816}. Cassandra has been described as a marriage of Dynamo and Bigtable \cite{Lakshman:2010:CDS:1773912.1773922}. Yahoo's PNUTS \cite{Cooper:2008:PYH:1454159.1454167} supports Yahoo! web properties such as Flickr. It uses per-record timeline consistency, where all replicas of a given record apply all updates to the record in the same order. Data objects (blobs) follow the eventually consistent model. 

Providing strong consistency guarantees, Microsoft's Windows Azure Storage (WAS)  \cite{Huang:2012:ECW:2342821.2342823} mixes strong consistency within a \emph{local stamp} (the analog of a local cluster for HCP) with subsequent replication to remote geographies. This model is the most similar to HCP, which provides strong write guarantees through synchronous replicas locally before asynchronously dispersing replicas geographically. Such systems might be referred to as \emph{eventually distributed}. 

Rather than requiring the \os{} to deal with collections of \fss{}, some implementations provide a single (logical) distributed \fs{}, such as the Google File System \cite{Ghemawat:2003:GFS:945445.945450} and CEPH \cite{Weil:2006:CSH:1298455.1298485}. By contrast, HCP uses a confederation of local file systems distributed over a collection nodes to persist data objects. This detail is abstracted from clients in favor of a storage service key-value model, where the key is a URI that indicates a unique object.

\section{Summary} \label{s:conclusion} 
In this paper we described \oss{} in general and HCP in particular, describing how they serve as a data container for \ud{}. We discussed the importance of \md{}, both to create logical collections of objects, and to provide the basis for establishing relation among objects. We then described an idealized relational distributed \os{} and described mechanisms for obtaining and applying \md{} to existing objects, and a method of associating and persisting relations amongst objects, providing a framework for data analytics to be performed against the \os{}.

\ifpdf
\pdfbookmark[1]{References}{references}
\fi

\small
\bibliographystyle{acm}        
\bibliography{master}          

\end{document}